\documentclass[9pt,twocolumn,twoside]{osajnl}

\journal{josab} 

\setboolean{shortarticle}{false} 

\title{Poor-man's model of hollow-core anti-resonant fibers}

\author[1,*]{Morten Bache}
\author[2]{Md. Selim Habib}
\author[1]{Christos Markos}
\author[1]{Jesper L\ae gsgaard}

\affil[1]{DTU Fotonik, Technical University of Denmark, Kgs. Lyngby, DK-2800, Denmark}
\affil[2]{CREOL, The College of Optics and Photonics, University of Central Florida, Orlando, FL-32816, USA}

\affil[*]{Corresponding author: moba@fotonik.dtu.dk}
\newcommand{\ts}{\Delta}
\newcommand{\neff}{n_{\rm eff, MS}}
\newcommand{\neffP}{n_{\rm eff, L}}
\newcommand{\neffPb}{\bar n_{\rm eff, L}}
\newcommand{\neffPt}{\tilde n_{\rm eff, L}}

\ociscodes{(190.4370) Nonlinear optics, fibers; (060.4005) Microstructured fibers; (060.2400) Fiber properties; (060.2310) Fiber optics}

\begin{abstract}
We investigate various methods for extending the simple analytical capillary model to describe the dispersion and loss of anti-resonant hollow-core fibers without the need of detailed finite-element simulations across the desired wavelength range. This poor-man's model can with a single fitting parameter quite accurately mimic dispersion and loss resonances and anti-resonances from full finite-element simulations. Due to the analytical basis of the model it is easy to explore variations in core size and cladding wall thickness, and should therefore provide a valuable tool for numerical simulations of the ultrafast nonlinear dynamics of gas-filled hollow-core fibers. 
\end{abstract}

\setboolean{displaycopyright}{true}

\begin{document}

\maketitle

\section{Introduction}

In the past decade broadband-guiding hollow-core fibers based on anti-resonant (AR) and inhibiting-coupling guiding mechanism, have become extremely popular (see recent reviews \cite{Wei2017.HCARF.review,Yu.HCARF.review.2016,markos-RevModPhys.89.045003}). These fibers support extremely large transmission bandwidths, making them excellent waveguides for studying ultrafast nonlinear optics in gases \cite{Travers:2011.JOSAB}. The only caveat is the presence of a number of sharp resonance bands, where the loss is very high and the dispersion is considerably affected. These resonances can simply be understood as a consequence of the presence of a thin glass capillary in the hollow fiber core boundary, but away from these resonance bands it turns out \cite{Im:2010} that the fiber dispersion is modeled excellently by the simple capillary model first outlined by  Marcatili and Schmeltzer \cite{Marcatili:1964}. 

Currently the accepted approach to model these fibers is to use this so-called Marcatili-Schmeltzer (MS) capillary model in nonlinear Schr\"odinger-like equations (NLSEs) and neglect the resonances and how they affect the dispersion and loss. Only recently did some of us include data from a full finite-element model (FEM) simulation of the dispersion and loss into the NLSE \cite{Habib2017,habib:2017.HCARF-UV-arxiv} and this was followed up recently by others where a Lorentzian extension of the MS model was implemented \cite{Tani2018,Sollapur2017}. 

Our motivation to find an analytical alternative to interpolating to FEM-based data is that the FEM simulations can be extremely cumbersome and difficult to do considering the very fine resolution needed to accurately locate the resonances. 
Especially our recent effort to understand how tapered hollow-core AR (HC-AR) fibers can be understood \cite{habib:2017.HCARF-UV-arxiv} underlined how difficult it was to provide enough FEM simulations for decreasing core sizes to really give an accurate picture of the losses and dispersion during the taper. We therefore decided to develop a poor-man's model of the FEM data, so that we could seamlessly track the resonances during the taper. 

A recent paper \cite{Zeisberger2017} showed a complete analytical description of dispersion and losses in thin capillary fibers. However, their perturbative approach was not able to capture the losses at the resonance wavelength. Ultimately, the original Marcatili and Schmeltzer paper \cite{Marcatili:1964} also addressed the extension of the basic MS model to include losses and higher-order dispersion contributions, but only for an infintely thick capillary. 
Our approach starts with a different non-perturbative approach, and from this we show how the perturbative approach can be modified to give accurate losses across the entire spectrum for a thin capillary, thus capturing both the  resonance and anti-resonance features, even in regimes where the glass cladding has significant losses. In turn, the resonant contributions to the mode dispersion are also accurately obtained. What remains to obtain the total loss of the HC-AR fiber is a single fitting parameter, expressing the deviation from a thin dielectric capillary to the more advanced HC-AR fiber designs emerging recently \cite{Belardi2014.HCARF.nested,Habib2016:ellipse,Habib:2015.nested,Poletti2014.HCARF.nested,Debord2017.HCARF.loss,Yu2016a.HCARF.experiment,Chaudhuri2016}. 

The idea is to calculate the eigenmodes of an HC-AR fiber, which typically has a core region (radius $a_c$) surrounded by some cladding elements that in some way or another defines the boundary region. The common element is that the core region is separated from the cladding region by a thin dielectric "wall" of thickness $\ts$. The core eigenmodes and their dispersion is easily calculated with the MS model, but the dispersion resonances and the losses are not contained in this analysis. To be able to model them, we take three different approaches. In the first case, we calculate the mode loss using a bouncing-ray model, and this we can use to calculate a Lorentzian extension of the dispersion using the knowledge that the loss spectrum has an associated resonance in the dispersion (Kramers-Kronig-like analogy). In the second case we perform a direct Kramers-Kronig transform of the found loss spectrum to find the dispersion. Finally, we show a perturbative extension of the MS model, which directly calculates the dispersion and loss resonances of the thin capillary by including the next-order contributions to the perfect-conductor case. 
We finally add to the loss the possibility to adjust for the details of the cladding structure (e.g., kagome, single-tube or nested-tube cladding elements etc.) and for material loss in the UV and mid-IR. 

\section{Capillary theory}

Let us first derive the equations of a hollow dielectric capillary, the so-called MS model \cite{Marcatili:1964}. 
Consider a dielectric capillary with radius $a_c$. In this simple model we do not consider the finite thickness of the capillary $\ts$; this will be taken into account later. The core supports a number of waveguide modes, each described by an effective index $\neff$. Formally we start from the relation $\beta^2=k_0^2-\kappa^2$ and introduce the effective index as $\beta=k_0\neff$. The transverse wavenumber of this eigenvalue is therefore
\begin{align}
\kappa=k_0 \sqrt{1-\neff^2}
\end{align}
where $k_0=\omega/c=2\pi/\lambda$ is the vacuum wavenumber. 
In the dielectric the transverse wavenumber is 
\begin{align}
\sigma=k_0\sqrt{n_d^2-\neff^2}
\end{align}
where $n_d$ is the dielectric refractive index. Since for a HC fiber $\neff\simeq 1$, irrespective of whether the fiber is evacuated or gas-filled, we can to a good approximation write $\sigma\simeq k_0\sqrt{n_d^2-1}$. 

We adopt the perfect-conductor approximation, which assumes that the guided core modes are zero at the core-dielectric interface. The radial nature of these modes are Bessel functions of the first kind $\propto J_m(\kappa r)$, yielding $J_m(\kappa a_c)=0$ under the perfect-conductor approximation. This implies that 
\begin{align}
\kappa a_c=u_{mn}
\end{align} 
where $u_{mn}$ is the $n$'th zero of the $m$'th order Bessel function $J_m$. 

The capillary model then predicts that effective mode index of the evacuated fiber can be calculated as $\neff^2=1-\kappa^2/k_0^2$ and under the perfect-conductor approximation we then get the effective index of the evacuated capillary in the MS model
\begin{align}
\label{eq:MS-dispersion}
\neff^2=1-\frac{u_{mn}^2}{a_c^2k_0^2}
\end{align}
This is the basis of the dispersion used in most of the simulations so far in the literature. 

However, this result does not capture the periodically resonant and anti-resonant behavior of the AR fiber because it does not take into account the finite thickness of the core-wall boundary $\ts$, i.e. a thin capillary  surrounded by vacuum both in the core and beyond the capillary. In this case it turns out that at certain resonance wavelengths distinct loss peaks appear, accompanied by resonance in the effective mode index (due to avoided-mode crossing between core and cladding modes), see e.g. \cite{Travers:2011.JOSAB}. We will below use 3 different approaches to modify the MS model to include the resonances by first calculating the resonant loss from a bouncing-ray approach. This is then used to include the dispersion resonances by extending the MS model with a series of associated Lorentzians at the resonant wavelengths. Then the same loss is used in an alternate approach where a Kramers-Kronig transform is used to calculate the dispersion. Finally, we show a perturbative extension of the MS model, where the dispersion and loss are directly calculated analytically by knowing the impedance response of the thin core-wall boundary. 

We should now mention briefly how to go from vacuum to a gas-filled fiber, although we will not use this in the rest of the paper where we will focus on the general vacuum case. As we shall see now, it is quite straight forward to generalize to a gas-filled fiber. Since the unity on the right-hand-side is reflecting the unity index of vacuum, we can for a gas-filled fiber replace this expression with the proper expression $n_{\rm gas}^2=1+\delta(\lambda)\tfrac{p T_0}{p_0 T}$, where $T$ is the gas temperature, $T_0=273.15$ K, $p$  is the gas pressure, $p_0=1$ atm, and $\delta(\lambda)$ models the wavelength dispersion of the gas. This corresponds to initially taking $\kappa=k_0 \sqrt{n_{\rm gas}^2-\neff^2}$ and we arrive at the celebrated expression of the MS model with gas dispersion included
\begin{align}
\neff=\sqrt{1+\delta(\lambda)\tfrac{p T_0}{p_0 T}-\frac{u_{mn}^2}{a_c^2k_0^2}}
\simeq 
1+\delta(\lambda)\tfrac{p T_0}{2p_0 T}-\frac{u_{mn}^2}{2a_c^2k_0^2}
\end{align}

\begin{figure}[t]
  \centering
  \includegraphics[width=\linewidth]{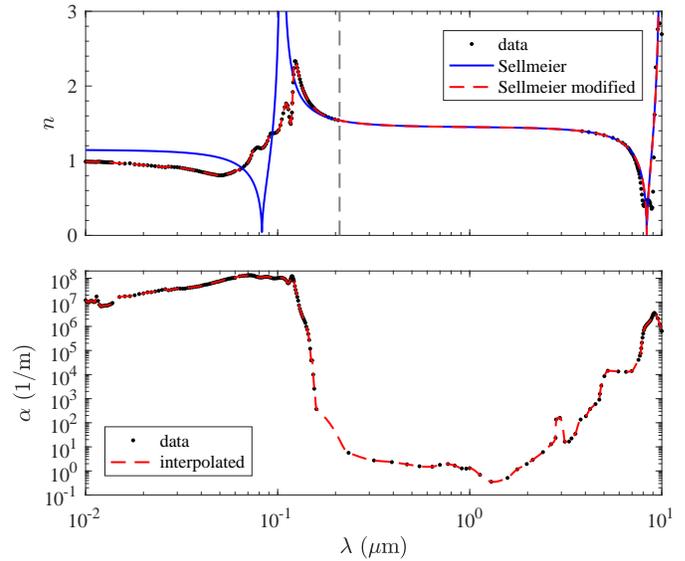}
\caption{Silica refractive index and loss coefficient vs. wavelength. The Sellmeier equation from \cite{ghosh:1994} is below 200 nm deviating from measured UV data \cite{palik:1998,Kitamura2007}. We therefore use a modified Sellmeier equation for $\lambda< 210$ nm (below dashed gray line) based on the UV data interpolated with a cubic spline. The dielectric loss coefficient $\alpha_d$ in Eq. (\ref{eq:alpha_pmm}) is also shown, based on interpolation to UV and IR loss data from \cite{palik:1998,Kitamura2007}. }
\label{fig:nSiO2}
\end{figure}

In order to calculate the analytical loss and dispersion using a silica HC-AR fiber, the challenge is that the wavelength range of interest in the community spans the extreme UV to the beginning of the mid-IR, essentially 50 nm-5.0 $\mu$m. For wavelengths shorter than 200 nm the silica Sellmeier equation we used \cite{ghosh:1994} is formally not valid, but by using available UV data (from \cite{palik:1998}, cf. also Fig. 2 in \cite{Kitamura2007}) we can avoid the spurious UV divergence in the Sellmeier model and get more accurate UV behavior. We therefore used the standard Sellmeier equation \cite{ghosh:1994} for $\lambda>210$ nm and the measured refractive index data of silica for $\lambda<210$ nm. In turn, the IR part of the Sellmeier equation fits well with IR data. Below we also need the material loss of silica across the range, also shown in Fig. \ref{fig:nSiO2}. 

\subsection{Bouncing-ray model of loss}

Following \cite{laegsgaard.2018} we now derive expressions for the modal loss (which should be mentioned is independent on gas pressure, unlike the modal dispersion) using a geometric "bouncing-ray" approach. Each mode will bounce between the capillary walls at a certain angle. We assume that the wall curvature is insignificant, which amounts to considering transmission and reflection between two dielectric sheets. The loss is found from calculating the transmission loss through the dielectric sheet during one bounce 
\begin{align}
\label{eq:alpha_BR}
\alpha_{\rm BR} \simeq (1-|r|^2)\frac{u_{mn}}{2a_c^2k_0}
\end{align}
Then by using Maxwell's equations and a forward-propagation traveling wave $e^{i(\beta z-\omega_0 t)} $ we can calculate the reflection $r$ and transmission $t$ amplitudes through the thin dielectric wall by matching the incoming and outgoing fields and their derivatives, and we get the following expressions 
\begin{align}
\label{eq:rTE}
r_{\rm TE}&=\frac{i(\frac{\kappa}{\sigma}-\frac{\sigma}{\kappa}) \tan(\sigma\ts)}{2+i(\frac{\kappa}{\sigma}+\frac{\sigma}{\kappa}) \tan(\sigma\ts)}
=\frac{\kappa-\frac{k_0}{Z_{\rm TE}}}{\kappa+\frac{k_0}{Z_{\rm TE}}}
\\
r_{\rm TM}&=
\frac{i(\frac{n_d^2\kappa}{\sigma}-\frac{\sigma}{n_d^2\kappa}) \tan(\sigma\ts)}{2+i(\frac{n_d^2\kappa}{\sigma}+\frac{\sigma}{n_d^2\kappa}) \tan(\sigma\ts)}
=
\frac{\kappa-\frac{k_0}{Y_{\rm TM}}}{\kappa+\frac{k_0}{Y_{\rm TM}}}
\label{eq:rTM}
\end{align}
where
\begin{align}
\label{eq:ZTE}
Z_{\rm TE}&=\frac{k_0}{\kappa} \frac{1-i\frac{\kappa}{\sigma} \tan(\sigma\ts)}
{1-i\frac{\sigma}{\kappa} \tan(\sigma \ts)}
\simeq 
Z_0\frac{1-i\frac{Z_d}{Z_0} \tan(k_0\ts/Z_d)}
{1-i\frac{Z_0}{Z_d} \tan(k_0\ts/Z_d)}
\\
\label{eq:YTM}
Y_{\rm TM}&=\frac{n_0^2 k_0}{\kappa} \frac{1-i\frac{n_d^2\kappa}{\sigma} \tan(\sigma\ts)}
{1-i\frac{\sigma}{n_d^2\kappa} \tan(\sigma\ts)}
\simeq
Y_0\frac{1-i\frac{Y_d}{Y_0} \tan(k_0\ts/Z_d)}
{1-i\frac{Y_0}{Y_d} \tan(k_0\ts/Z_d)}
\end{align}
Eq. (\ref{eq:ZTE}) was also found in \cite{Miyagi:1984}.
We have here introduced 
\begin{align}
Z_d&=(n_d^2-1)^{-1/2}\simeq k_0/\sigma\\
Y_d&=n_d^2(n_d^2-1)^{-1/2}\simeq k_0n_d^2/\sigma
\end{align}
being the dielectric surface impedance and admittance, respectively. We have also introduced the impedance of the mode in the material surrounding the dielectric (here vacuum, $n_0=1$)
\begin{align}
Z_0&=k_0/\kappa=k_0a_c/u_{mn}\\
Y_0&=n_0^2k_0/\kappa=Z_0=k_0a_c/u_{mn}
\end{align}
where the last identity in each equation holds in the perfect conductor approximation that underlies the whole analysis. We will throughout the paper focus on the ratios $\sigma/\kappa\simeq Z_0/Z_d$ and $\sigma/(n_d^2\kappa)\simeq Y_0/Y_d$. 


Using the above equations we arrive at the power loss coefficients
\begin{align}
\label{eq:alphaTE}
\alpha_{\rm TE, BR}&=\frac{2u_{mn}}{a_c^2k_0 [
4\cos^2(\sigma \ts)+(\tfrac{\kappa}{\sigma}+\tfrac{\sigma}{\kappa})^2 \sin^2(\sigma\ts)]}
\\
\label{eq:alphaTM}
\alpha_{\rm TM,BR}&=\frac{2u_{mn}}{a_c^2k_0 [
4\cos^2(\sigma \ts)+(\tfrac{n_d^2\kappa}{\sigma}+\tfrac{\sigma}{n_d^2\kappa})^2 \sin^2(\sigma \ts)]}
\end{align}
For hybrid modes, including the fundamental mode with $m=0$ and found by taking the first zero $n=1$, the loss is taken as a geometric average 
\begin{align}
\alpha_{\rm H,BR}&=(\alpha_{\rm TE,BR}+\alpha_{\rm TM,BR})/2
\label{eqn:alpha_hybrid}
\end{align}
Depending on what mode we consider, the relevant capillary loss will then be TE, TM or hybrid. 

We should here note that the loss in what follows is calculated from Eqs. (\ref{eq:alphaTE})-(\ref{eqn:alpha_hybrid}) using $n_d$ real; consequently $\sigma$ and $Z_0$ are taken real. The latter is not given in the deep UV as $n_d<1$ can be seen, but on the other hand then the HC-AR fiber no longer supports guided modes so we restrict our analysis to the guided-mode regime. If the losses should be evaluated from the complex refractive index of silica (i.e. taking into account the absorption coefficient) then it must be stressed that silica in the UV is not an ideal metal, but rather a lossy dielectric. This dissipative system makes the boundary conditions more complicated, cf. e.g. \cite{Carson:1936}, as the perfect conductor assumption cannot be taken. Later we will show how the expressions can be analytically generalized to include a lossy dielectric in the cladding, based on results from \cite{Archambault:1993}. 

From these expressions, in the limit of large core sizes $a_ck_0\gg 1$ we have $Z_0\gg Z_d$, so it becomes clear that the loss is minimized whenever $\sigma \ts=(2l+1)\pi/2$, where $l$ is an integer; this is the famous anti-resonance condition for low-loss guidance. In turn, the resonance condition for maximum loss is fulfilled when $\sigma \ts=l\pi$. In the following, the resonant wavelengths are therefore found by solving the resonance condition 
\begin{align}
\label{eq:lambdaR}
\lambda_R=\frac{2\ts\sqrt{n_d^2(\lambda_R)-1}}{l}, \quad l=1,2,...
\end{align}
Note that the resonance conditions are the same for TE, TM and hybrid modes, and it is only the resonance width that changes: in the TE case it is a factor $n_d^2$ narrower than the TM case.

The minimal loss ("valleys" of loss spectra at the anti-resonance wavelength) can be found by taking $\sin(\sigma\ts)=1$. At the resonances, instead we have $\sin(\sigma\ts)=0$.  
The extremes of the hybrid mode loss therefore become 
\begin{align}
\label{eq:alpha_min}
\alpha_{\rm H, BR}^{\rm min}&\simeq 
\frac{u_{mn}^3(Z_d^2+Y_d^2)}{a_c^4 k_0^3}=\frac{u_{mn}^3(n_d^4+1)}{a_c^4 k_0^3 (n_d^2-1) }
\\
\label{eq:alpha_max}
\alpha_{\rm H, BR}^{\rm max}
&=\frac{u_{mn}}{2a_c^2 k_0}
\end{align}
where the minimum loss holds in the limit $Z_0\gg Z_d$.

\subsection{Lorentzian extension of dispersion}

We now show one way of modeling the dispersion resonances from knowing the loss. This takes a Drude-like approach, where the relative permittivity $\varepsilon=n^2$ is modified from the MS expression by adding Lorentzian resonances  
\begin{align}\label{eq:neff_Lorentzian}
\neffPb^2(\omega)=\neff^2(\omega)+\sum_{l=1}^{n_{\rm res}} \frac{B_{R,l}\omega_{R,l}^2}{\omega_{R,l}^2-\omega^2-i\omega\Gamma_{R,l}}
\end{align}
Here $n_{\rm res}$ is the total number of resonances included, $\Gamma_{R,l}$ is the resonance width and $B_{R,l}$ the resonance strength. Note that with this definition we get a complex effective index $\neffPb=\neffP+i\neffPt$. We will only use the real part $\neffP$ for the model, while the imaginary part will be used to extract information about the resonances as we will now see, using that $2k_0\neffPt$ is equivalent to the loss parameter $\alpha$. 

We note that this approach is not the same as used in \cite{Sollapur2017,Tani2018}, where similar Lorentzian shapes were added to $n$ instead of $n^2$. This gives subtle but noticeable differences in the linewidth shape, and we argue that the shape we use from a physical standpoint is sounder.

The resonance strengths $B_{R,l}$ can be calculated by the following argument: assuming the resonances are not too close, the loss at the resonance wavelength can be directly related to the imaginary part of the complex refractive index at the resonance $\alpha(\omega_{R,l})=\neffPt(\omega_{R,l})2\omega_{R,l}/c$. At the same time, we find $\neffPt(\omega_{R,l})\simeq B_{R,l}\omega_{R,l}/[2 \neff(\omega_{R,l})\Gamma_{R,l}]$, assuming that the real part $\neffP$ can be approximately described by the capillary model at the resonance. This is as we shall see a good approximation since the resonances turn out not to be too wide. Combining these two expressions we get that the $l$'th resonance strength can be calculated as
\begin{align}
\label{eq:BR}
B_{R,l}=\frac{c\alpha_{\rm H,BR}(\omega_{R,l}) \neff(\omega_{R,l})\Gamma_{R,l}}{\omega_{R,l}^2}
\end{align}

\begin{figure}[t]
  \centering
  \includegraphics[width=\linewidth]{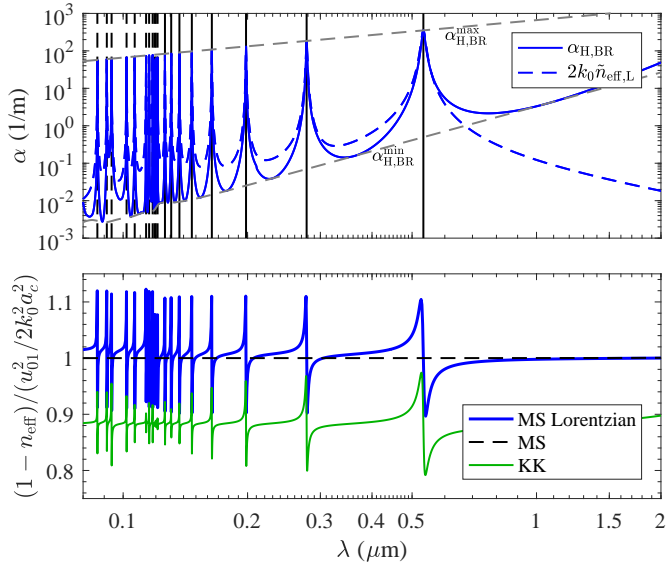}
\caption{Top plot: Bouncing-ray calculation of the loss vs. wavelength (log scale) of the fundamental hybrid HE$_{01}$ mode in a 17 $\mu$m core radius evacuated silica capillary with thickness $\Delta=250$ nm. The loss was calculated with Eqs. (\ref{eq:alphaTE})-(\ref{eqn:alpha_hybrid}), and the Lorentzian loss $2k_0\neffPt$ was calculated from Eq. (\ref{eq:neff_Lorentzian}) using $n_{\rm res}=21$. Of these are 8 major resonance wavelengths (black lines) that were numerically calculated from Eq. (\ref{eq:lambdaR}) with a root-finding method. The 13 secondary resonance wavelengths (black dashed lines) correspond to zeros found below the peak of silica refractive index at 123 nm. The gray dashed curves show the scaling of the minimum and maximum loss in the bouncing-ray model. The scaled effective index (bottom plot) was in the MS model calculated from Eq. (\ref{eq:MS-dispersion}) and in the poor-man's model the extensions Eqs. (\ref{eq:neff_Lorentzian})-(\ref{eq:GammaR}) were applied. The result of a Kramers-Kronig transformation, Eq. (\ref{eq:n_KK}), is also shown, based on the bouncing-ray loss $\alpha_{\rm H,BR}$. For the Kramers-Kronig transformation $2^{13}$ equidistant angular frequency points were used in the range $1-30,000$ THz.}
\label{fig:loss-noFEM}
\end{figure}

The linewidth of the resonances $\Gamma_{R,l}$ were found by matching the spectral loss shapes  calculated from the imaginary part of the effective index from the Lorentzian oscillator model with the one from the hybrid loss. The principle behind this exercise is demonstrated in Fig. \ref{fig:loss-noFEM}: the capillary loss of the fundamental hybrid mode $m=0$ and $n=1$ (HE$_{01}$) was calculated with Eqs. (\ref{eq:alphaTE})-(\ref{eqn:alpha_hybrid}). The first task in our central algorithm is to numerically locate the major resonance wavelengths from Eq. (\ref{eq:lambdaR}) for $l=1,2,\dots$ (black lines), and we remark that they agree perfectly with the loss peaks present in $\alpha_{\rm H,BR}$; we found this task numerically easier than the alternative approach of locating the maxima of $\alpha_{\rm H,BR}$. Since the refractive index of silica is not monotonous towards the UV, see Fig. \ref{fig:nSiO2}, we may find for each $l$ value a number of additional zeros from Eq. (\ref{eq:lambdaR}); these are marked as black dashed lines and are added also to the model. In other words, $n_{\rm res}$ is not just the chosen maximum number of $l$ but the secondary resonances are added as well. Next, the dispersion of the poor-man's model is calculated by first matching the Lorentzian resonance strengths $B_{R,l}$ so the loss represented by $2k_0\neffPt$ equals that of $\alpha_{\rm H,BR}$ at the resonance wavelengths, which is done by using Eq. (\ref{eq:BR}). Finally, the Lorentzian linewidths are adjusted so the losses match as well as possible not just at the resonance wavelengths but also in the valleys. The following empirical relationship turned out to be very useful
\begin{align}
\label{eq:GammaR}
\Gamma_{R,l}=\frac{\omega_{R,1}}{40 l}
\end{align}
It is not presently clear why this expression works so well, but it seems to be general to all the cases we have tested. We also used this expression when fixing the linewidths of the higher-order zeros due to the decreasing UV refractive index. Judging by Fig. \ref{fig:loss-noFEM} the dashed line, representing the loss calculated from the Lorentzian extension of the capillary dispersion, matches very well the loss calculated by the ray-tracing approach. This validates the approach. 

We do note that the IR behavior of the loss is not modeled in the Lorentzian case, which can be amended by adding an IR loss term to Eq. (\ref{eq:neff_Lorentzian}). This is not so important in the poor-man's model model per se, since we will in any case use Eq. (\ref{eqn:alpha_hybrid}) for modeling the waveguide loss. Also the loss in the valleys is somewhat larger in the Lorentzian case, which might be because the TE and TM modes have different linewidths and this is not taken into account in the Lorentzian model (where only a single linewidth is used at each resonance).  

A remarkable feature of the higher-order resonance conditions in the UV is that both due to the increasing material refractive index in the UV and due to the increased order, the distance between successive resonances becomes very small, especially considering the logarithmic scale used on the wavelength axis. Even for this ideal case where no material loss is included the AR fiber has practically no transmission bands in the UV. This should motivate a design with thinner capillary walls so the lowest antiresonance valley is blue-shifted away from the pump wavelength and where the UV resonances will be more distant. Later we will see that when taking into account the lossy nature of the cladding dielectric in the UV, these loss peaks will be smeared out.

Fig. \ref{fig:loss-noFEM} also shows how the modal dispersion is affected by the Lorentzian lines we add to the capillary model: the resonances are seen as sharp jumps in the effective index at the resonance wavelengths, compared to the smooth behavior of the MS model. The effective index scaling we use in Fig. \ref{fig:loss-noFEM} is chosen from Eq. (\ref{eq:MS-dispersion}), where we approximately get that $\neff\simeq 1 -u_{mn}^2/(2a_c^2k_0^2)$, so that in the MS model we get that the dispersion of the vacuum case is unity when the scaling is chosen as $(1 -\neff)/[u_{mn}^2/(2a_c^2k_0^2)]$. 

\subsection{Kramers-Kronig transformation}

An alternative approach to derive the dispersion from a known  loss spectrum is to invoke a Kramers-Kronig transform, which connects the real and imaginary parts of the linear susceptibility $\chi=\chi'+i\chi''$ as
\begin{align}
\chi'(\omega)&=\frac{1}{\pi}\mathcal{P}\int_{-\infty}^{\infty} d\omega'\frac{\chi''(\omega')}{\omega'-\omega}
\\
&=\frac{2}{\pi}\mathcal{P}\int_{0}^{\infty} d\omega'\frac{\omega'\chi''(\omega')}{\omega'^2-\omega^2}
\end{align}
where $\mathcal{P}$ indicates the Cauchy principal value. The second stage was derived using the fact that the time response is real, thus providing a deterministic link between positive and negative frequencies. Introducing now the connection with the complex refractive index $1+\chi=\bar n^2$ and using the weakly absorbing approximation we get $\bar n=n+i\tilde n\simeq 1+\chi'/2+i\chi''/2$, which finally gives the expression we look for
\begin{align}
\label{eq:n_KK}
n(\omega)=1+\frac{c}{\pi}\mathcal{P}\int_{0}^{\infty} d\omega'\frac{\alpha(\omega')}{\omega'^2-\omega^2}
\end{align}
where we used that $\alpha(\omega)=\omega\chi''(\omega)/c$. 
The transformation was done both with an available Matlab script \cite{KK-matlab} that uses basic linear integration, as well as with a home-written script using trapezoidal integration rules.

In Fig. \ref{fig:loss-noFEM} we also show the Kramers-Kronig transformation results. This used the hybrid mode loss of the bouncing-ray model to calculate the effective mode index. We observe an excellent agreement between the resonances strengths between the two approaches. The advantage of the Kramers-Kronig transform is that we do not make any educated or empirical guesses to the linewidths and the strengths, so the overall agreement actually confirms that we did a pretty good job in determining the Lorentzian parameters. However, clearly the Kramers-Kronig result has an offset. 
This offset turns out to be very sensitive to the boundaries  of the numerical transform (in principle we should integrate from 0 to $\infty$) and also whenever the frequency resolution was altered the offset changed dramatically. This seems to be an inherent problem with the Kramers-Kronig transform: it must integrated well beyond a resonance, and on top of that $\chi'(\omega)$ must converge to zero faster than $1/\omega$, which it turns out not to do in the specific case. Finally, the transformation can become very slow even for rather modest array sizes. This often means that calculating the transform can be the same computational order as the numerical integration of the NLSE itself. We can therefore not recommend this approach.  

\subsection{Perturbative extension of the MS model}

We finally show a third way of calculating the loss and directly obtain the associated dispersion resonances. It relies on introducing a complex solution to the propagation constant (i.e. eigenvalue) to first order. We take our starting point in the work of \cite{Miyagi:1984,Marcatili:1964} who to first order calculate the complex eigenvalues in the MS model, i.e. the ideal capillary case. The starting point is in \cite{Marcatili:1964} where for the infinitely thick capillary the transverse core wave number $\kappa$ was extended beyond lowest order as 
\begin{align}
\kappa a_c=
   \begin{cases}
        u_{mn}(1-i\frac{Z_d}{k_0a_c}), & \text{TE}\\
      	u_{mn}(1-i\frac{Y_d}{k_0a_c}), & \text{TM}
    \end{cases}
\end{align}
The correction comes from the fact that the true eigenfunction solution relies on Hankel functions, that to lowest order can be considered constant, which gives the basic perfect-conductor result, but to first order has a correction $\mathcal{O}[(k_0a_c)^{-1}]$ \cite{Marcatili:1964}. 
Using this extension the propagation constant becomes complex, i.e. we allow now a complex effective mode index. 
This holds for $Z_d\ll Z_0$. 
The implications are directly found as follows (TE case)
\begin{align}
\beta&=k_0\sqrt{1-\kappa^2/k_0^2 }
\simeq k_0\left [1-\frac{u_{mn}^2}{2a_c^2k_0^2}\left(1-\frac{i2Z_d}{a_ck_0}\right)\right]
\end{align}
which can straightforwardly modified to obtain the TM case. From this equation one can calculate the mode loss from the imaginary part, while the real part contains the dispersion. 

We now observe the calculated TE and TM reflection coefficients for the thin capillary, Eqs. (\ref{eq:rTE})-(\ref{eq:rTM}). When the capillary is thick the reflection coefficient for the TE mode becomes $ r_{\rm TE}= (\kappa-\frac{k_0}{Z_{\rm d}})/(\kappa+\frac{k_0}{Z_{\rm d}})$. In other words, when going from an infinite to a finite, thin capillary, $Z_{\rm TE}$ can be viewed as the generalized impedance needed to calculate the reflection coefficient, replacing the dielectric impedance $Z_d$ in the infinite case. In the same spirit, we can generalize the perturbative extension of $\kappa$ to the thin capillary case, i.e. where the dielectric impedance $Z_d$ and admittance $Y_d$ are replaced by the TE and TM results Eqs. (\ref{eq:ZTE})-(\ref{eq:YTM}). This gives the following perturbative extension of the effective index and power loss coefficient for a hybrid mode
\begin{align}
\label{eq:neff_p}
n_{\rm eff,p} &= \neff-\frac{u_{mn}^2}{a_c^3k_0^3}{\rm Im}[\tfrac{1}{2}(Z_{\rm TE}+Y_{\rm TM})]
\\
\label{eq:alpha_p}
\alpha_{\rm H,p}&=\frac{2u_{mn}^2}{a_c^3k_0^2}{\rm Re}[\tfrac{1}{2}(Z_{\rm TE}+Y_{\rm TM})]
\end{align}
Since the impedance has a periodic recurrence of maxima and minima, as dictated by the resonance and anti-resonance conditions, this is a successful first-order extension of the MS model to include the resonances in dispersion and loss, remarkably without any additional assumptions. A similar expression as Eq. (\ref{eq:alpha_p}) was found in \cite{Miyagi:1984}. We should mention that we have discarded higher order terms in the impedances for simplicity, but it is easy to keep them in regimes where they might give relevant corrections. 

\begin{figure}[t]
  \centering
  \includegraphics[width=0.9\linewidth]{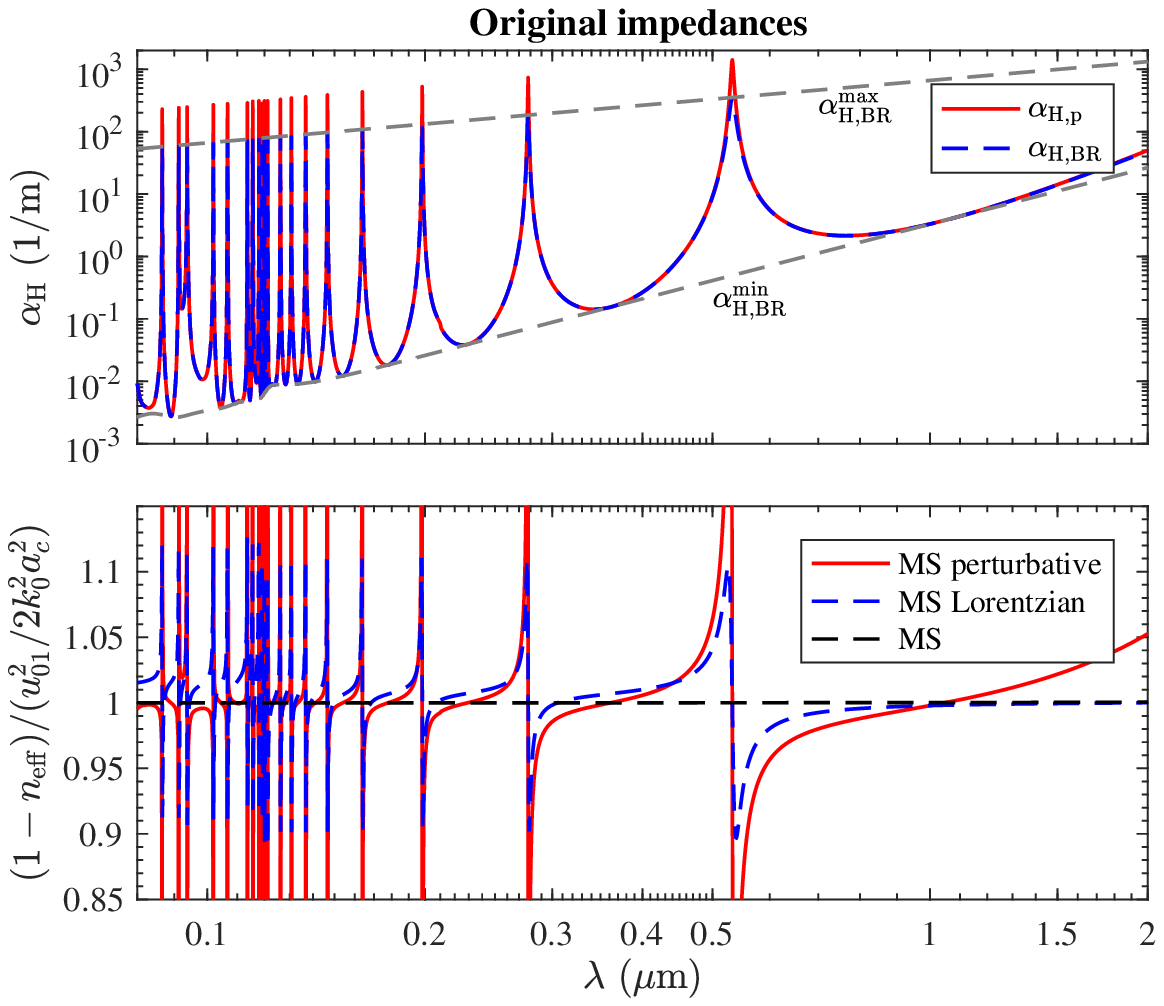}
  \includegraphics[width=0.9\linewidth]{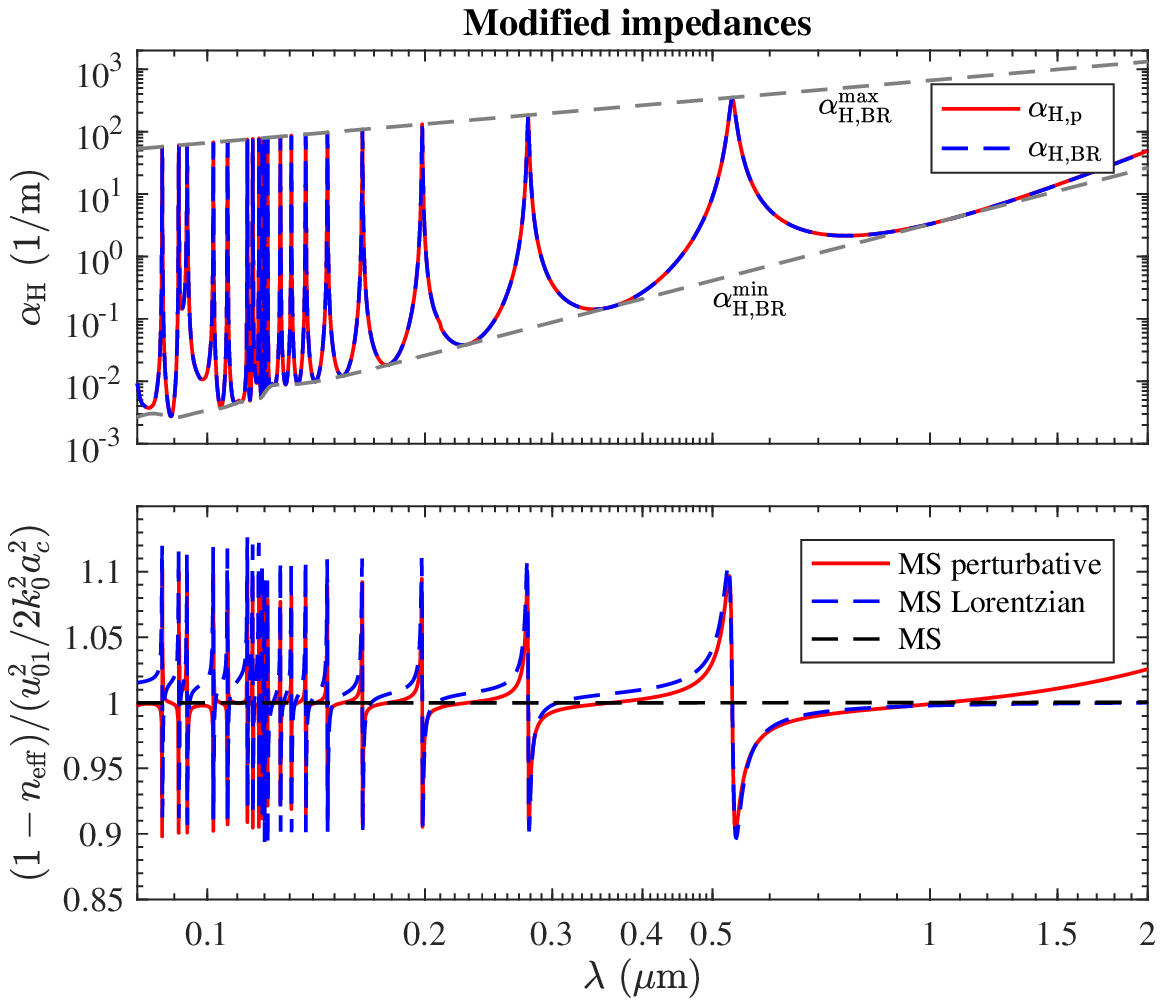}
\caption{As Fig. \ref{fig:loss-noFEM}, but using the perturbative extension of the MS model Eqs. (\ref{eq:neff_p})-(\ref{eq:alpha_p}). Top plots: using the original impedances Eqs. (\ref{eq:ZTE})-(\ref{eq:YTM}). Bottom plots: using the empirically modified impedances Eqs. (\ref{eq:ZTE_mod})-(\ref{eq:YTM_mod}). The MS Lorentzian curves are the same as in Fig. \ref{fig:loss-noFEM}. }
\label{fig:loss-noFEM-perturbative}
\end{figure}

In Fig. \ref{fig:loss-noFEM-perturbative} we show the calculations of the loss and the dispersion using the perturbative extension to the MS model. The hybrid loss is very similar to the bouncing-ray result in Fig. \ref{fig:loss-noFEM}. In fact, a direct comparison shows that the they match perfectly over the entire range. Only the peak values are higher by exactly a factor 4. It is worth to mention that at resonance wavelength the requirement $Z_{TE}\ll Z_0$ is not fulfilled; in fact the value is precisely $Z_{TE}=Z_0$ at resonance. The perturbative treatment therefore breaks down, which was also discussed in \cite{Archambault:1993}. In fact, they specifically show in their Fig. 7 that at resonance the analytical expression for the loss, which is identical to Eq. (\ref{eq:alpha_p}), is significantly larger than numerical simulations of the loss. We also compared the bouncing-ray loss values for the same geometry they used i Fig. 7 and achieved good agreement with the numerical simulation results close to the resonance. This seems reasonable as the bouncing-ray model has no specific limitations at resonance.  
In fact, what happens is that exactly at the resonance we get $r_{\rm TE/TM}=0$, so that $|t_{\rm TE/TM}|^2=1$ and we immediately get $\alpha_{\rm TE/TM}=u_{mn}/(2a_c^2 k_0)$ from Eq. (\ref{eq:alpha_BR}). We are therefore inclined to trust the bouncing-ray model result close to and at resonance. 

Remarkably, if the impedances are empirically modified as follows
\begin{align}
\label{eq:ZTE_mod}
\hat Z_{\rm TE}=Z_0\frac{\frac{1}{2}-i(\frac{\sigma}{\kappa}+\frac{\kappa}{\sigma})^{-1} \tan(\sigma \ts)}
{2-i(\frac{\sigma}{\kappa}+\frac{\kappa}{\sigma}) \tan(\sigma \ts)}
\\
\label{eq:YTM_mod}
\hat Y_{\rm TM}=Y_0\frac{\frac{1}{2}-i(\frac{\sigma}{n_d^2\kappa}+\frac{n_d^2\kappa}{\sigma})^{-1} \tan(\sigma \ts)}
{2-i(\frac{\sigma}{n_d^2\kappa}+\frac{n_d^2\kappa}{\sigma}) \tan(\sigma \ts)}
\end{align}
and we then use these in Eq. (\ref{eq:alpha_p}) to calculate the loss, we obtain exactly the bouncing-ray loss Eq. (\ref{eq:alphaTE})-(\ref{eq:alphaTM}). This is  demonstrated in Fig. \ref{fig:loss-noFEM-perturbative}. 
We should here stress that if we in the bouncing-ray model use the impedance analogy, e.g. $r_{\rm TE}=(\kappa-\frac{k_0}{Z_{\rm TE}})/(\kappa+\frac{k_0}{Z_{\rm TE}})$, to calculate the loss, then Eqs. (\ref{eq:alphaTE})-(\ref{eq:alphaTM}) would of course also be modified once we change the impedances, but as can be seen in the first expressions in Eq. (\ref{eq:rTE})-(\ref{eq:rTM}) we can arrive at the bouncing-ray losses without resorting to the impedance analogy. 
These modified impedances are important because they are instrumental in obtaining the analytical dispersion extension, which then will give a more correct dispersion at the resonance; remember from the Lorentzian case that the peak loss determines the resonance strength, cf. Eq. (\ref{eq:BR}). In fact, when comparing the dispersion from the modified impedances to the original ones, we see that the line shapes are much less sharp, and actually very similar to the ones derived with the Lorentzian extension. This is exactly due to the reduced losses in the resonances. 

\begin{figure}[t]
  \centering
  \includegraphics[width=0.9\linewidth]{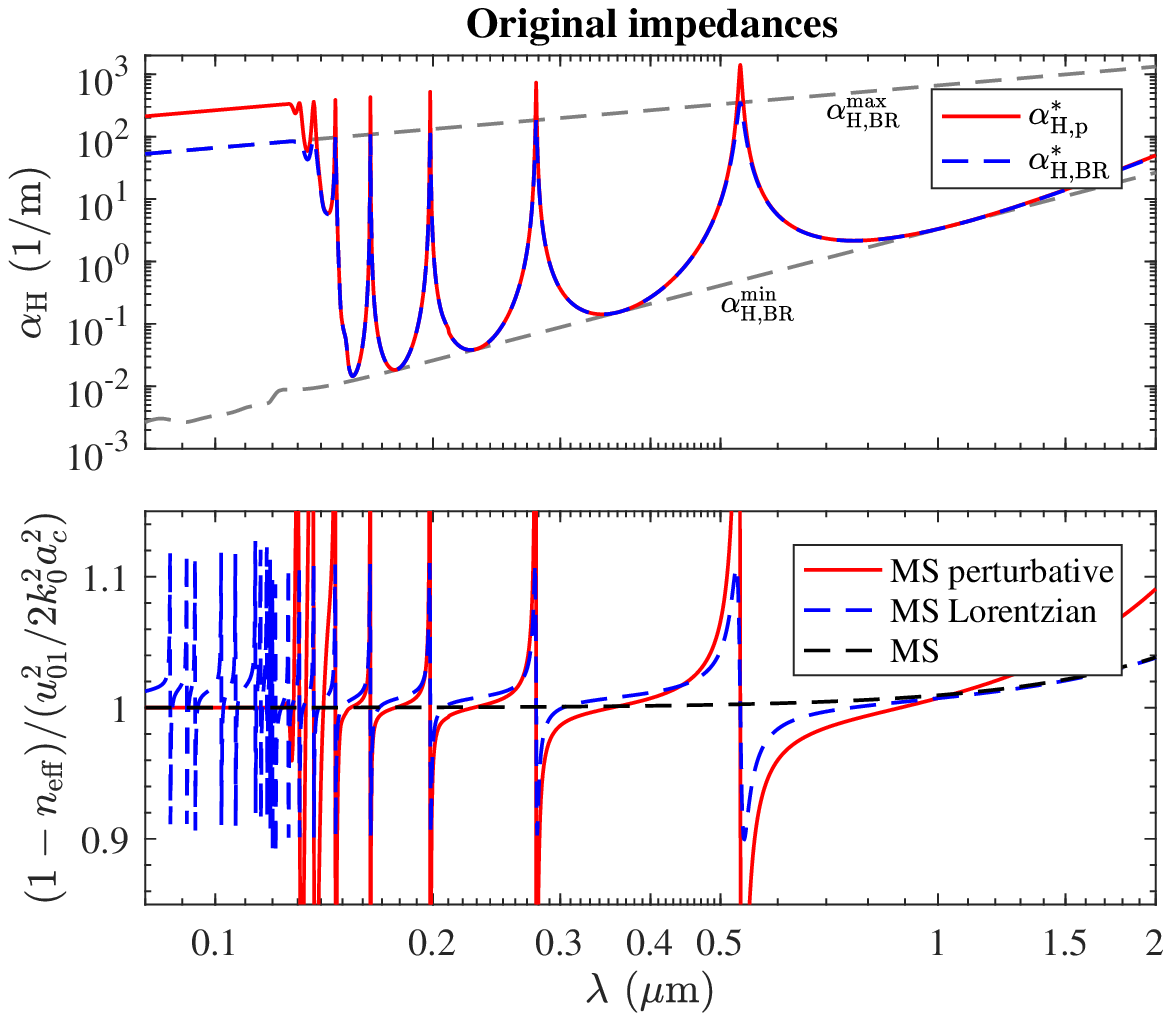}
  \includegraphics[width=0.9\linewidth]{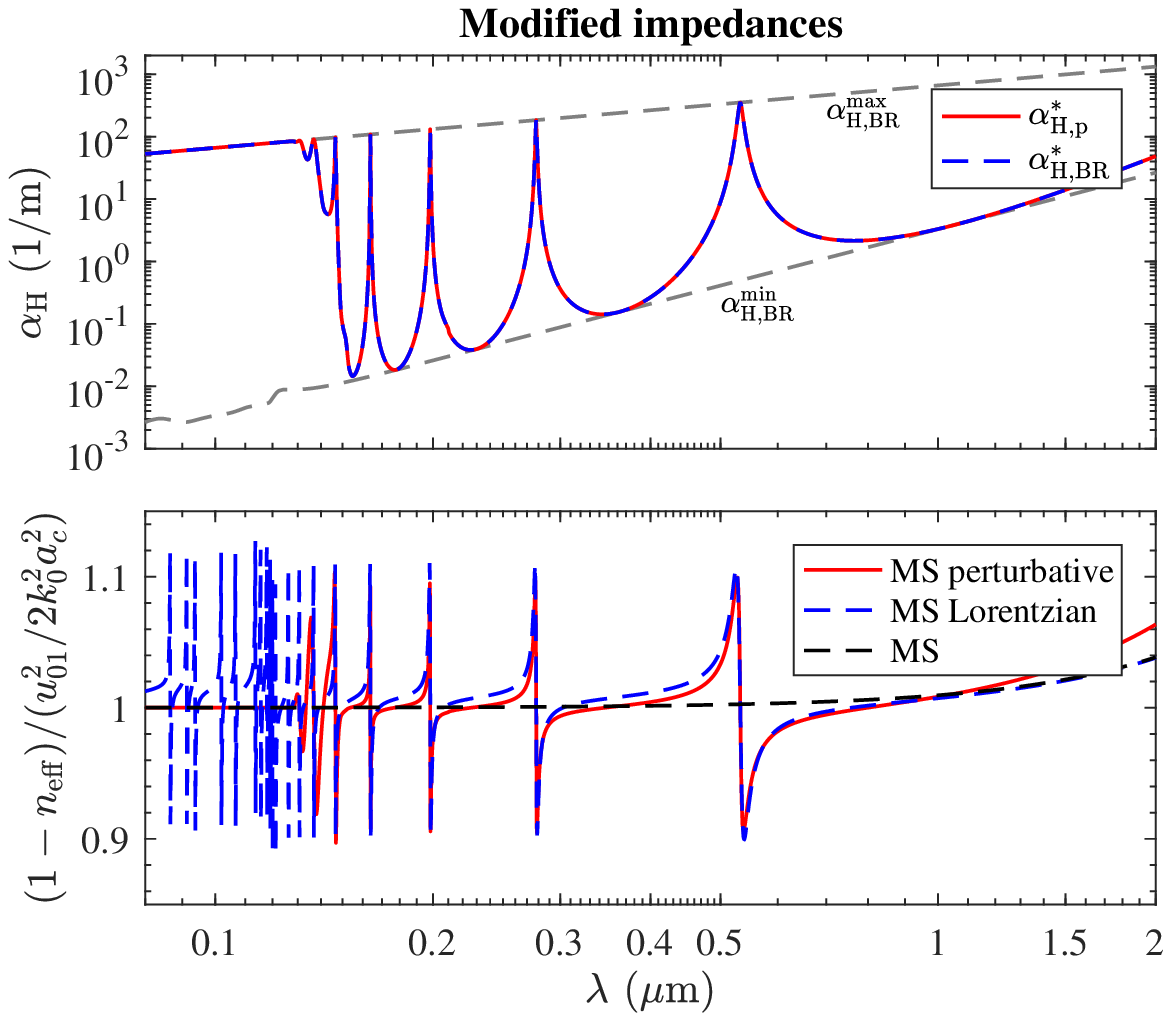}
\caption{As Fig. \ref{fig:loss-noFEM-perturbative}, but taking into account the lossy nature of silica in the UV, 
i.e. using Eqs. (\ref{eq:sigmakappa_lossy_TE})-(\ref{eq:sigmakappa_lossy_TM}). The MS Lorentzian dispersion curves are the same as in Fig. \ref{fig:loss-noFEM}, while the bouncing-ray losses were taken from Eqs. (\ref{eq:alphaTE})-(\ref{eq:alphaTM}) and substituting in Eqs. (\ref{eq:sigmakappa_lossy_TE})-(\ref{eq:sigmakappa_lossy_TM}). 
}
\label{fig:loss-noFEM-perturbative-lossy}
\end{figure}

Interestingly, in \cite{Archambault:1993} they also discuss how the perturbative losses can be calculated if the  dielectric is lossy $\bar n_d=n_d+i\tilde n_d$. It turns out that the we must replace the transverse wavenumber ratio $\sigma/\kappa$ 
for the TE case with
\begin{align}
\label{eq:sigmakappa_lossy_TE}
\left(\frac{\sigma}{\kappa}\right)^*=
\frac{\sigma}{\kappa} \frac{1+\frac{\kappa}{\sigma}\tanh(n_d \tilde n_d Z_d^2\sigma\ts )}{1+\frac{\sigma}{\kappa}\tanh(n_d \tilde n_d Z_d^2\sigma\ts )}
\end{align}
Similarly for the TM case 
we must replace $\sigma/(n_d^2\kappa)$ with
\begin{align}
\label{eq:sigmakappa_lossy_TM}
\left(\frac{\sigma}{n_d^2\kappa}\right)^*=
\frac{\sigma}{n_d^2\kappa} \frac{1+\frac{n_d^2\kappa}{\sigma}\tanh(n_d \tilde n_d Z_d^2\sigma\ts )}{1+\frac{\sigma}{n_d^2\kappa}\tanh(n_d \tilde n_d Z_d^2\sigma\ts )}
\end{align}

With these extensions we can evaluate analytically the expected mode loss due to the UV material loss in silica. This is done in Fig. \ref{fig:loss-noFEM-perturbative-lossy}, which essentially takes the same approach as Fig. \ref{fig:loss-noFEM-perturbative}. The perturbative loss, here denoted with a $^*$ superscript to indicate that it is using the lossy silica extension, is now heavily modified in the UV. In fact, with the original impedances the loss becomes totally dominated by material losses below 120 nm; this is because in the limit of large $\tilde n_d$ we have $(\sigma/\kappa)^*\rightarrow 1$. This has profound consequences for the losses: for the TE case in this limit ${\rm Re}(Z_{\rm TE})=Z_0$, i.e. constant, and no resonances appear. 
In the same way, we see that the dispersion resonances also vanish, which is a consequence of ${\rm Im}(Z_{\rm TE})=0$ in this limit and therefore the dispersion equals that of the MS model. Similar arguments hold for the TM case. However, as before the loss peaks are overestimated compared to the bouncing-ray model. This is again solved by using the modified impedances. It is worth noting that by using 
Eqs. (\ref{eq:sigmakappa_lossy_TE})-(\ref{eq:sigmakappa_lossy_TM}) in 
the bouncing-ray loss expressions Eqs. (\ref{eq:alphaTE})-(\ref{eq:alphaTM}), we also obtain the heavily modified UV losses. We note that the Lorentzian extension of the dispersion is not easily modified based on this new loss profile, because the resonance wavelengths are calculated assuming that $Z_0\gg Z_d$, equivalent to $\sigma/\kappa\gg 1$. However, this is no longer true when we use $(\sigma/\kappa)^*$, which as we discussed above becomes unity when $\tilde n_d$ becomes significant. Therefore the Lorentzian extension falsely predicts a number of dispersion resonances in the UV.



Finally we remark that the analytical model in \cite{Zeisberger2017} is also relying on a perturbative extension of the MS model, but it cannot predict the absolute loss at the resonances since the loss was found to be $\propto \cot(\sigma \Delta)$, and therefore it becomes infinity exactly at the resonance. 

We find that the perturbative extension is by far the most powerful since it has no basic assumptions about the lineshape, linewidth and strengths, and it can be instantaneously calculated once the impedances are properly defined. 

\subsection{Total loss}

In what follows we compare the analytical loss and dispersion calculations with FEM data. We have two approaches: the lossy dielectric and the ideal dielectric cases. 

\begin{figure}[t]
  \centering
  \includegraphics[width=\linewidth]{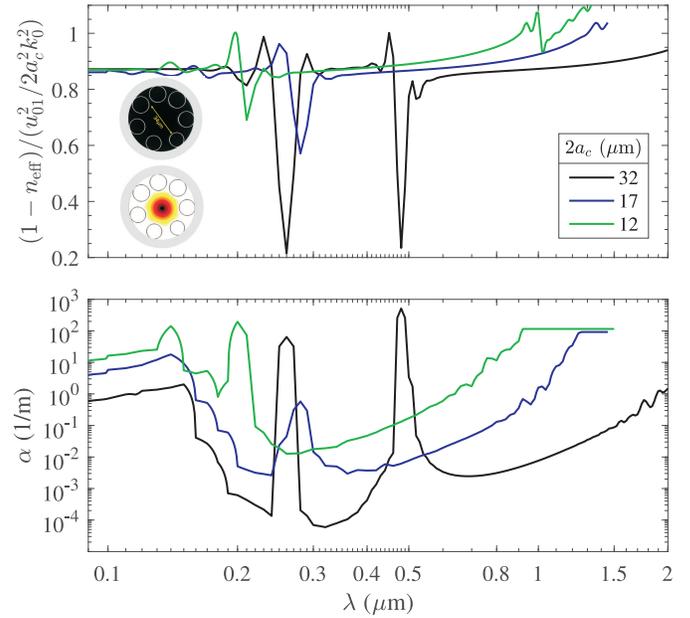}
\caption{FEM data of calculated effective index and loss vs. wavelength, originally used in \cite{habib:2017.HCARF-UV-arxiv}. A SEM image of a fabricated fiber, see inset, was used as a starting point, and the shown cores sizes are then considered linearly tapered down so $\Delta$ is scaled accordingly down. The effective index scaling is chosen from Eq. (\ref{eq:MS-dispersion}). 
Inset: SEM image of a fabricated single-ring HC-AR silica fiber with 7 AR cladding tubes,  ($2a_c=34~\mu$m and $\Delta=250$ nm) 250 nm tube-wall thickness and $34~\mu$m core diameter. The FEM design used in this paper based on this SEM image, and the fundamental mode at 800 nm is shown from the FEM calculation. }
\label{fig:fiber_loss}
\end{figure}

In the lossy dielectric case we have an analytical prediction of the total mode loss across the spectrum by using the complex refractive index of silica. In this case we therefore model the total loss as
\begin{align}\label{eq:alpha_pmm_lossy}
\alpha_{\rm total}^* =f_{\rm FEM}\alpha_{\rm capillary}^*
\end{align}
$f_{\rm FEM}$ is the overall fitting factor that allows us to adjust the capillary spectral loss shape to match the levels found in COMSOL. This is important because while it turns out that the AR fiber loss shape can be accurately predicted this way, we cannot predict the absolute loss values because this depends on the particular overall AR fiber design. Importantly, this factor seems to be common to a particular design: we successfully used the same value for a range of COMSOL simulations all based on the same fiber design (i.e. the AR fiber with single-ring tube cladding AR elements, cf. Fig. \ref{fig:fiber_loss}). 

Instead in the ideal dielectric case the total loss in the poor-man's model is now calculated based on the above capillary expressions for the TE, TM and hybrid modes as follows
\begin{align}\label{eq:alpha_pmm}
\alpha_{\rm total} =f_{\rm FEM}\alpha_{\rm capillary}+\alpha_d F_d
\end{align}
We have here introduced some parameters that allow us to include material loss in high-absorption regions like the UV and mid-IR, namely $\alpha_d$ (the dielectric loss coefficient), and $F_d$ (the power fraction of light residing in the dielectric). In what follows we will seek to get some insight into the scaling laws governing the $F_d$ parameter. 

It turns out that the power fraction of light residing in the dielectric has a non-trivial scaling relation to (1) the wavelength, and (2) the core radius. A clue to how this might behave can be found from considering the analytical loss from the bouncing-ray model, cf. Eqs. (\ref{eq:alpha_min})-(\ref{eq:alpha_max}) . 

What we are mainly interested in incorporating into the model is an overall material loss, and we therefore focus on the behavior away from the resonances; around the resonances we will have a separate (and in most cases dominant) contribution to the overall loss from the resonance peaks in the calculated modal resonant losses. 

In order to conduct a quantitative study, we need to consider detailed FEM simulations. The FEM data were calculated from the UV to the near-IR for a range of core sizes and cladding tube thicknesses, all on an evacuated fiber ($p=0$). 
A scanning electron microscope (SEM) image of the single-ring HC-AR fiber used in our calculations is shown in Fig. \ref{fig:fiber_loss} (inset). The fabricated HC-AR fiber has a core diameter of $2a_c=34~\mu$m, an average capillary diameter of $16~\mu$m, and an average silica wall thickness of $\Delta=250$ nm. The  wall thickness was chosen to give a first AR transmission band centered at 800 nm. The near-field profile of the fundamental mode of the imported cross-section structure, calculated using FEM, is also shown.  
We used the refractive index and material loss of silica from Fig. \ref{fig:nSiO2} (essentially we invoked a complex refractive index of silica) to calculate the mode propagation constant $\beta$ and confinement loss using the FEM-based COMSOL software. This approach is different than our previous work where the silica loss was not considered in the FEM calculation and the material-based loss was added afterwards based on the power-fraction of light in silica \cite{Habib:2015.nested}. 
To get an accurate calculation of the loss, we used a perfectly-matched layer outside the fiber domain and great care was taken to optimize both mesh size and the parameters of the perfectly-matched layer \cite{Habib:2015.nested,Poletti2014.HCARF.nested,habib:2016.semicircle}. 

Examples of the FEM data for the scaled effective mode index and the confinement loss are shown in Fig. \ref{fig:fiber_loss} for 3 selected fiber sizes, where the original fiber design is assumed to be linearly scaled down to the selected core size, so this means that $\Delta$ scales accordingly as well. We see that the refractive index resonances are also associated with large loss peaks; the first resonance is the strongest, and a second resonance can sometimes be located as well. It is, however, problematic to find the resonances in the UV region due to the large material loss of silica, which is responsible for the loss edge starting at 200 nm. 
The deviation from unity of the scaled refractive index in the UV range is related to the fact that $\pi a_c^2$ underestimates the core area; therefore it is customary to use a modified MS (mMS) model where a generalized wavelength-dependent core area is used in the MS model $a_c(\lambda)=a_{\rm AP}/[1+s\lambda^2/(a_{\rm AP}\Delta) ]$ \cite{Finger2014}. Here 
an "area-preserving" core radius $a_{\rm AP}$ is introduced, which can be chosen to compensate for the UV asymptotic behavior to match the FEM data. In the IR we note that the dispersion also deviates from unity. Traditionally, this was addressed in the mMS model by introducing the $s$ parameter. The specific values we chose to use in the mMS model are detailed below when we eventually compare the theory with the FEM data.

\begin{figure}[t]
  \centering
  \includegraphics[width=\linewidth]{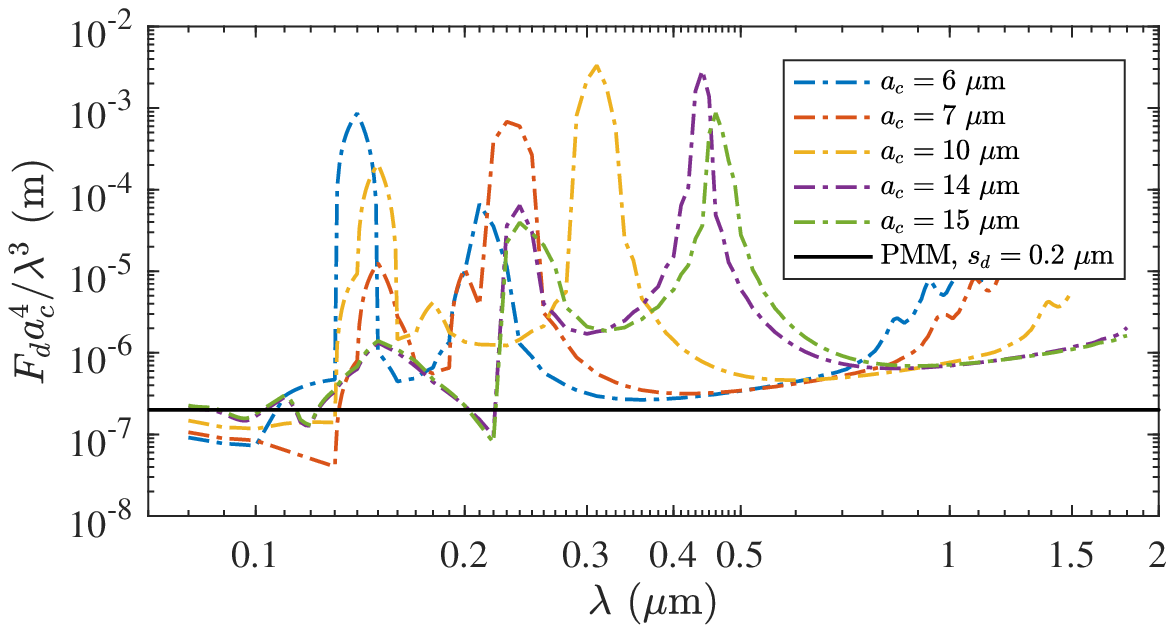}
  \includegraphics[width=\linewidth]{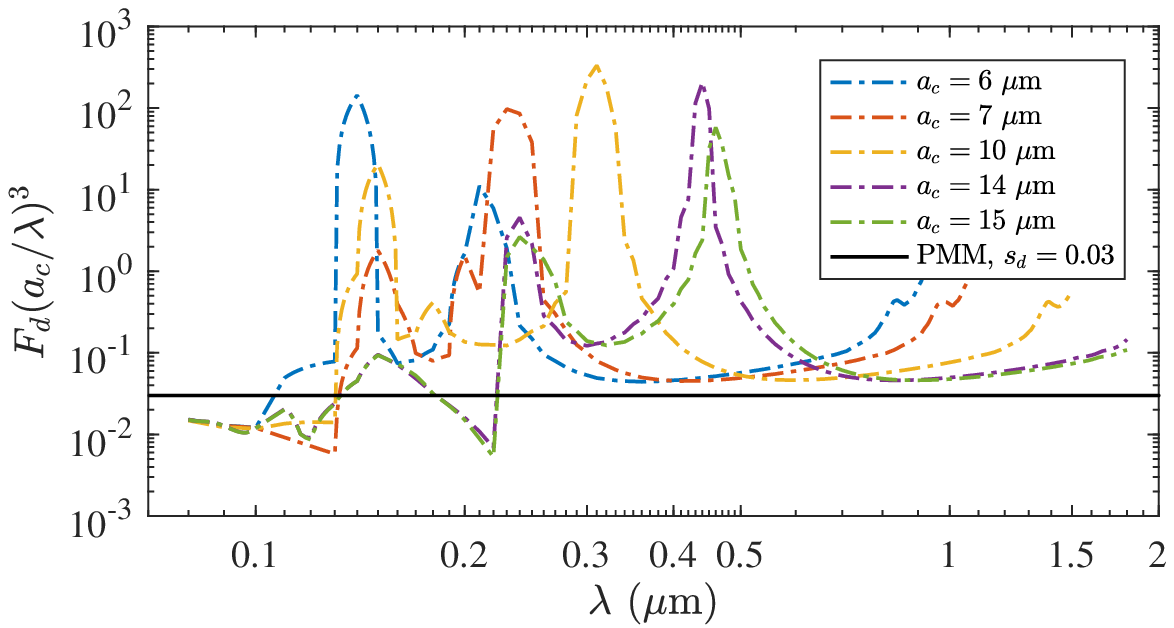}
  \includegraphics[width=\linewidth]{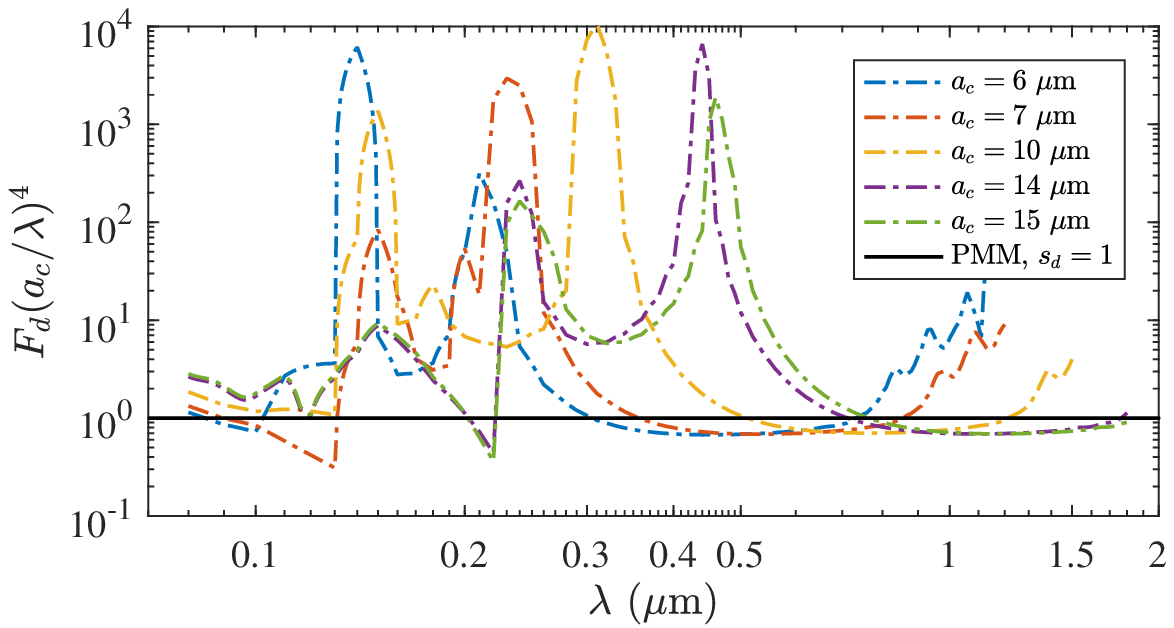}
\caption{Fraction of power in dielectric (silica) vs. wavelength. The FEM data are taken for 5 different core radii, and the values are normalized to the proposed scaling laws $\lambda^3/a_c^4$ (top), $\lambda^3/a_c^3$ (middle) and $\lambda^4/a_c^4$ (bottom). }
\label{fig:fops}
\end{figure}
\begin{figure*}[t]
  \centering
  \includegraphics[width=0.48\linewidth]{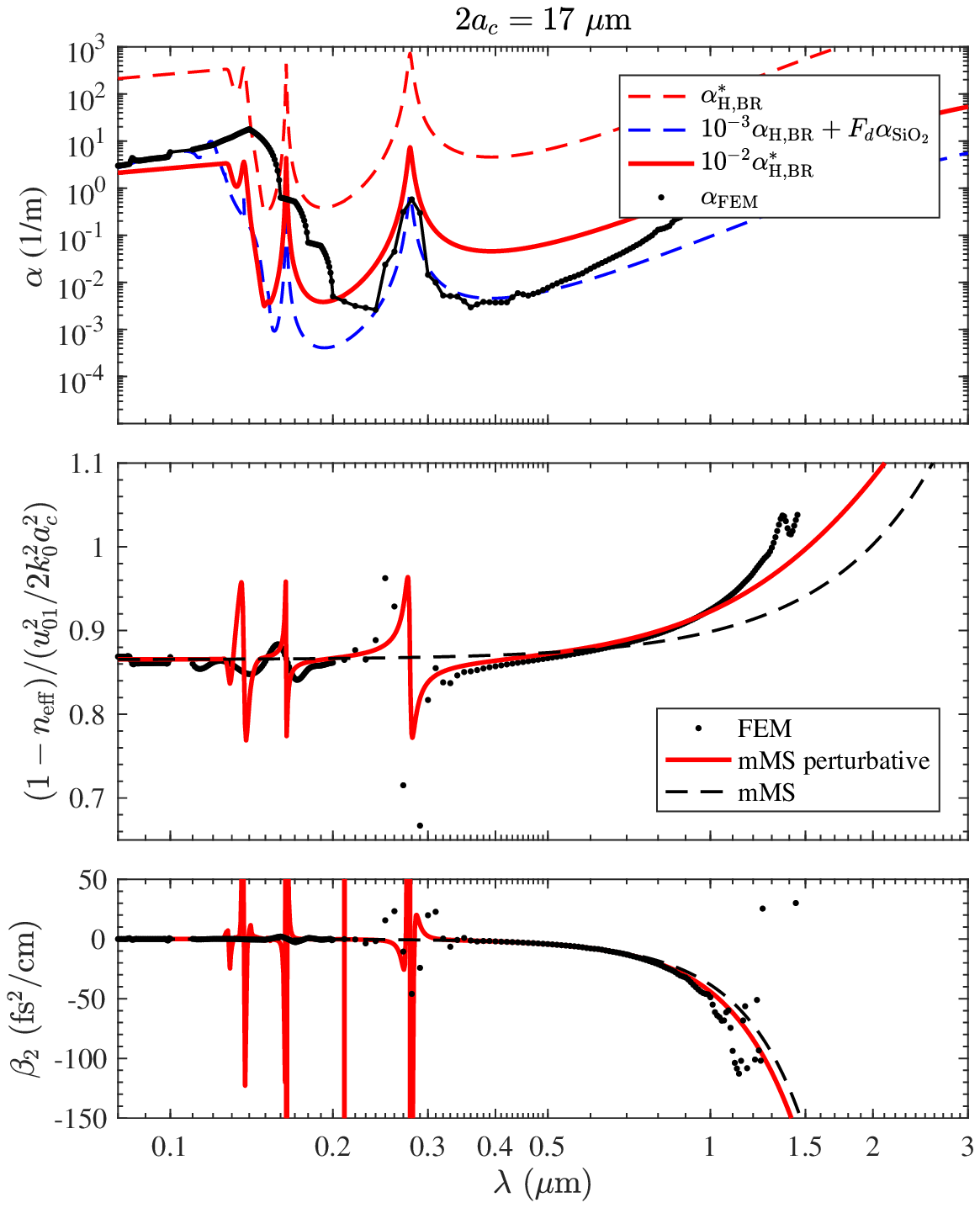}
  \includegraphics[width=0.48\linewidth]{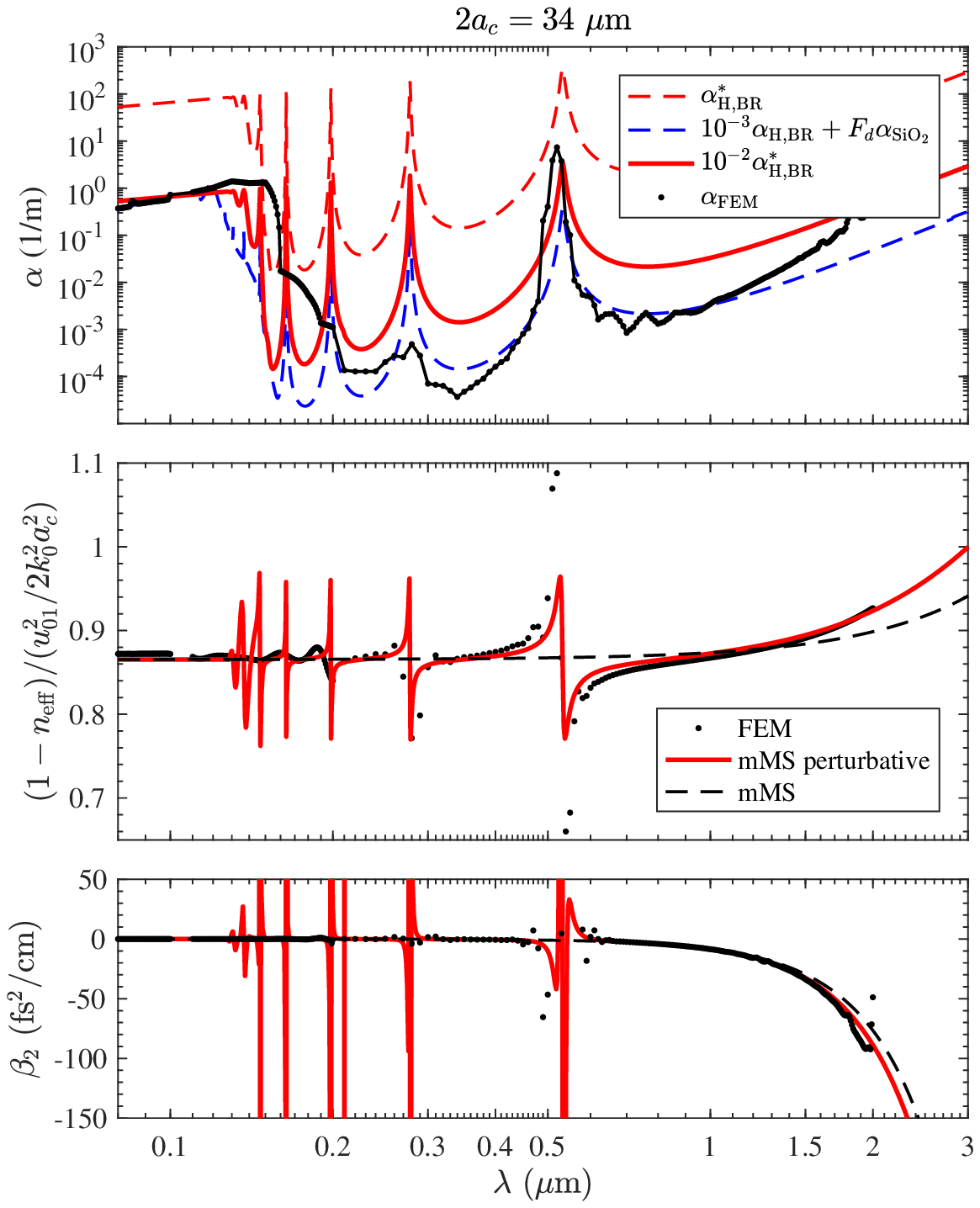}
\caption{As Fig. \ref{fig:loss-noFEM-perturbative-lossy} plotted together with FEM data. The total loss based on the lossy silica calculation $\alpha_{\rm total}^*$ from Eq. (\ref{eq:alpha_pmm}), using $f_{\rm FEM}=10^{-2}$ to match the FEM data. The total loss based on the loss-less silica case is also shown, where the glass loss is modeled by including the dielectric (silica) material loss extracted from \cite{Kitamura2007}, and $\alpha_{\rm total}$ from Eq. (\ref{eq:alpha_pmm}) using $f_{\rm FEM}=10^{-3}$ and $F_d=0.03(\lambda/a_c)^3$. We note that in this figure the mMS model with the parameters $s=0.02$ and $a_{\rm AP}=1.075a_c$. Left: $2a_c=17~\mu$m and $\ts=125$ nm, right: $2a_c=34~\mu$m and $\ts=250$ nm. The FEM simulations were taken from the same simulations behind Fig. \ref{fig:fiber_loss} and \ref{fig:fops}, and are originally from \cite{habib:2017.HCARF-UV-arxiv}. 
}
\label{fig:loss-FEM}
\end{figure*}

As argued above, the hybrid mode losses away from the resonances scale as $\alpha_{\rm H, BR}^{\rm min}\propto 1/(a_c^4 k_0^3)\propto \lambda^3/a_c^4$. The proposed scaling law $F_d=s_d\lambda^3/a_c^4$ is shown in Fig. \ref{fig:fops} (top) where $s_d$ is a scaling coefficient. We also tested two other simple scaling laws $F_d=s_d(\lambda/a_c)^m$ where $m$ an integer. All cases seem to predict quite well the $F_d$ FEM data values; especially the visible and near-IR ranges are described very well. The resonant wavelengths are seen to carry a much higher power fraction in glass, and this is quite impossible to model as it depends on how the cladding modes look like. We also note that the XUV parts of the spectra do not overlap for $m=4$; we actually see that the XUV increase that is seen below 100-130 nm seems to scale as $\lambda^3/a_c^3$. Why this is so must be investigated further, but choosing this scaling leaves also the choice of the scaling coefficient: $s_d=0.03$ slightly underestimates the value in the visible, but gives an accurate level in the far- and XUV (see also later in Fig. \ref{fig:loss-FEM}). Arguably, it is the latter regime that is important because it is exactly there that the UV losses of silica kick in.  

We are now finally  ready to compare the the analytical expressions of the total loss with FEM data. In the lossy dielectric case, all we need to find is the FEM correction factor by comparing with FEM data. This is done in Fig. \ref{fig:loss-FEM} for two selected core sizes. Here we have the dilemma that the analytical calculation cannot match both the anti-resonant valleys and the UV plateau. Choosing $f_{\rm FEM}=10^{-2}$ gives a suitable compromise. 
This correction factor is very fiber-dependent as, e.g., very intricate designs of the AR elements can give extremely low losses \cite{Belardi2014.HCARF.nested,Habib2016:ellipse,Habib:2015.nested,Poletti2014.HCARF.nested,Debord2017.HCARF.loss,Yu2016a.HCARF.experiment,Chaudhuri2016}. 
We note an overall good agreement between the FEM case and the analytical curves  in the loss spectra, and also the resonances in the refractive indices are well reproduced. The main discrepancy with the FEM data is that the UV loss edge sets in earlier in the FEM data. 

In the ideal, lossless dielectric case we need to  combine the scaling of the power fraction in glass, cf. Fig. \ref{fig:fops}, and the material loss vs. wavelength, cf. Fig. \ref{fig:nSiO2}. 
In order to adjust the overall loss level we now address the FEM correction factor, cf. Eq. (\ref{eq:alpha_pmm}). We found that $f_{\rm FEM}=10^{-3}$ gave good agreement with the COMSOL data, see examples of two test cases in Fig. \ref{fig:loss-FEM}. This dual-parameter fit gives an overall better agreement with the FEM data than in the lossy dielectric case. It is therefore not entirely obvious which method to use, and more detailed studies are needed to compare the analytical model to more ideal FEM test cases. We are inclined to recommend using the perturbative method for a lossy dielectric with the modified impedances, which with a single stroke and one single fitting parameter gives both the loss and dispersion in any wavelength range as long as the complex refractive index data of the dielectric is available. 


As mentioned above we use the mMS model \cite{Finger2014} in order to obtain a correction in the IR and UV dispersion. 
The area-preserving radius can be determined using the UV asymptotic level of the normalized effective index in Fig. \ref{fig:loss-FEM} (it is responsible for shifting the mMS data below unity), and we note that we find a higher value ($a_{\rm AP}=1.075 a_c$) than in a kagome fiber (where $a_{\rm AP}=a_c\sqrt{2\sqrt{3}/\pi}\simeq 1.05a_c$ was found \cite{Finger2014}). 

Concerning the requirement for an IR dispersion correction 
the perturbative case shows an good agreement in the IR with the FEM data for $s=0.02$. This is somewhat lower than previous studies and is a consequence of the perturbative extension of the dispersion gives additional IR contributions compared to the ideal MS dispersion, cf. Eq. (\ref{eq:neff_p}). 

\section{Conclusion}
In conclusion 
we have tested various analytical extensions of the capillary model to describe the anti-resonant and resonant transmission bands and how they affect the dispersion as well as loss. 
Two of these methods were based on calculating the loss spectrum in a general, non-perturbative way, and then associating this loss with an imaginary part of the refractive index. Of these, the Lorentzian model was the most successful, while for the Kramers-Kronig transform of the loss spectrum it was difficult to obtain good convergence. The third one relied on a perturbative extension of the basic capillary model, i.e. taking into account a complex refractive index to first order beyond the perfect conductor approximation. This approach is not accurate at the resonance wavelengths, but we showed how to modify it to obtain perfect agreement with the non-perturbative loss calculations. This method was ultimately the most successful since it does not rely on any assumptions (unlike the Lorentzian case) and in the modified form we present here the losses as well as the dispersion should be very accurate across the resonance and anti-resonance bands. 
Importantly, our model also takes into account the material loss of the dielectric in the cladding, which is important in the UV for silica-based fibers. The model relies only on a single overall fitting parameter, which needs to be determined with a few test finite-element simulations for a given fiber design to give the overall total loss (to take into account intricate cladding designs intended for ultra-low loss fibers). 
We compared to FEM data and found that considering the simplicity the overall agreement is quite impressive. This analytical extension of the capillary model is a quick way of mimicking complicated and detailed finite-element simulations, so we expect it to find broad usage in the community.

\section{Acknowledgements}
We thank Ole Bang for useful discussions. Enrique Antonio-Lopez and Rodrigo Amezcua Correa are acknowledged for drawing the HC-AR fiber we based our design studies on. 

\bibliography{literature}
\end{document}